\documentclass[aps,prd,superscriptaddress,amsmath,amssymb,twocolumn,preprint,nofootinbib,10pt]{revtex4-2}

\usepackage{graphicx} 
\usepackage{dcolumn}
\usepackage{bm}
\usepackage[bookmarks=false]{hyperref}
\usepackage{color}

\definecolor{mygreen}{RGB}{0,155,0}

\def\c#1{\textcolor{black}{#1}}

\begin{document}

\title{Spin induction from scattering of two spinning black holes in dense clusters}
\author{Jorge L. Rodríguez-Monteverde}\email[]{jorge.lopezr@estudiante.uam.es}
\affiliation{Universidad Autónoma de Madrid, Cantoblanco 28049 Madrid, Spain}

\author{Santiago Jaraba}\email[]{santiago.jaraba@uam.es}
\author{Juan García-Bellido}\email[]{juan.garciabellido@uam.es}
\affiliation{Instituto de Física Teórica UAM-CSIC, Universidad Autónoma de Madrid, Cantoblanco 28049 Madrid, Spain}

\date{\today}

\begin{abstract}
In this paper, we \c{use} numerical relativity \c{to study} the spin induction effect within close hyperbolic encounters of \c{initially spinning} black holes. We review the initially non-spinning case and explore the cases of initially aligned, anti-aligned and orthogonal spins with respect to the orbital angular momentum $\vec L$. We find that, for a given initial effective spin, the black hole with a smaller initial spin acquires a greater spin-up than the other black hole after the \c{interaction}. We study three different scenarios regarding initial effective spin ($\chi_{\rm eff}=-0.1, \ 0.0, \ 0.1$), using three different scattering angles in order to obtain maximally spin-inducing scenarios. We also find that the final effective spin\c{-ups} with respect to the initial spins are well-fitted by a parabola. \c{For} spins orthogonal to $\vec L$, we observe that the black hole spins precess, and that the induced spin in the $z$-direction depends quadratically on the value of the initial spins. These phenomena suggest that dense black hole clusters \c{present a rich spin dynamics}, where black hole spins may acquire non-trivial distributions.
\end{abstract}

\preprint{IFT-UAM/CSIC-2024-147}

\maketitle
\section{Introduction}
\label{Sec:Intro}
In the last decades, remarkable progress has been made in order to characterize compact binary coalescences (CBC) in the most accurate and complete way possible. These efforts have only been enhanced since the first detection of a gravitational wave (GW) in 2015~\cite{LIGOScientific:2016aoc}, originated by a binary black hole (BH) system. However, as the GW detectors become increasingly sensitive, other signals of different nature may be observed. This is the case of hyperbolic encounters: unbound two-body interactions in which both masses scatter off each other. In recent years, these encounters have been studied from different approaches, usually related to their GW emission, from their general formalism~\cite{Garcia-Bellido:2017qal,Garcia-Bellido:2017knh,Teuscher:2024xft} and more detailed modelling~\cite{Caldarola:2023ipo,Roskill:2023bmd,AbhishekChowdhuri:2023rfv,Fontbute:2024amb} to their detectability with current~\cite{Morras:2021atg,Bini:2023gaj} and future~\cite{Mukherjee:2020hnm,Kerachian:2023gsa,Barrau:2024kcb} GW detectors. \c{The abundance and phenomenology of these interactions closely depend on the specific properties of black hole populations. This is particularly interesting from the perspective of primordial black holes (PBH) in dense clusters~\cite{Clesse:2016vqa,Clesse:2016ajp,Garcia-Bellido:2017fdg,Trashorras:2020mwn,Siles:2024yym}, where BH scattering is common. The detailed study of these events and their detection with GW observatories will allow us to test whether some fraction of these compact objects lies within dense clusters, as usually considered in the thermal history scenario~\cite{Carr:2019kxo,Carr:2023tpt}. Moreover, if these events are frequent enough, they would produce a stochastic gravitational wave background with abundant information on the BH clustering properties~\cite{Garcia-Bellido:2021jlq}.}

\c{Instead}, we focus on another relevant feature of the dynamics of hyperbolic encounters: the spin induction effect. Previous works using Numerical Relativity (NR)~\cite{Nelson:2019czq,Jaraba:2021ces} showed that, when these black hole scatterings have small impact parameter, also known as close hyperbolic encounters (CHE), they are able to induce significant spins on initially non-spinning black holes. \c{In addition, more recent works~\cite{Chiaramello:2024unv} aim to accurately model this effect analytically}. A good characterization of \c{spin induction} is key to understanding the evolution of the spin distribution in black hole clusters, which should be accounted for in future N-body simulations~\cite{Siles:2024yym}.

Both of these works on spin induction assume initially non-spinning black holes, either for equal masses~\cite{Nelson:2019czq} or introducing mass asymmetries~\cite{Jaraba:2021ces}. In the PBH scenario, this is a reasonable assumption due to their expected low spins. However, in dense clusters where the spin induction effect is relevant, we expect the initial spins of these encounters to be increasingly higher as they evolve in time. In addition, works as~\cite{Rettegno:2023ghr} show that initial spins have a noticeable impact on the scattering angles, which are tightly related to the final BH spins. It is thus necessary to study the effects of initial spins on the spin induction.

In this work, we explore the consequences of the introduction of initial non-zero spins. For this purpose, we run a series of NR simulations, obtaining the dependence of the final spins in terms of the initial ones. We will consider spins both parallel and orthogonal to the orbital angular momentum, the latter producing spin precession\footnote{\c{Although spin precession was originally introduced for binary systems having shorter orbital period than the spin precession timescale~\cite{PhysRevD.49.6274,Gerosa_2023}, we have found this terminology convenient to describe the spin evolution in CHEs, since the BH spins suffer a change in the orientation of the rotation axis during this motion.}} and impacting the three spin components of each BH. Finally, we will compare our results with simple analytical expressions so as to explain the dependence we obtain in terms of the initial spin for the final BH spins. This more ambitious exploration of the parameter space in terms of the initial spins needs, however, to reduce the scope in other parameters such as the mass ratio. Therefore, we only consider equal-mass binaries in this article, while the exploration of the simultaneous effect of unequal masses and initial spins is left for future work. Additionally, we will not consider the dependence of this effect with the incidence angle, as it is done in~\cite{Nelson:2019czq,Jaraba:2021ces}. 

This work is structured as follows. In Sec.~\ref{Sec: Theory}, we provide some theoretical context for a better comprehension of the results. \c{Then, we describe the configuration used in our NR simulations in Sec.~\ref{Sec: Grid}}. Later, in Sec.~\ref{Sec: Results}, we describe our numerical results for equally spinning black holes, while the unequally spinning case is presented in Sec.~\ref{Sec: Numerical results 2}, together with a discussion of the results. We conclude with some final remarks in Sec.~\ref{Sec: Concl}.

The results presented in this work are described using geometrized units, $G=c=1$.

\section{\c{Black hole interaction}}
\label{Sec: Theory}
\c{Before presenting the numerical simulations, we describe how two BHs interact, for a better understanding of the results. We characterize the interaction between two spinning BHs with the following Hamiltonian:}
\begin{equation}
    H\simeq H_{\rm orb}+H_{SO}+H_{S_1S_2},
    \label{eq: hamiltonian}
\end{equation}
\begin{equation}
    \c{H_{\rm SO}\simeq \frac{2G}{c^2r^3}\vec{L}\cdot\vec{S}_{\rm eff}},
    \label{eq: spin-orbit}
\end{equation}
\begin{equation}
   \c{ H_{S_1S_2}\simeq-\frac{G}{c^2r^3}\Big(\vec{S}_1\cdot\vec{S}_2-3(\vec{S}_1\cdot \hat{\vec r}_1)(\vec{S}_2\cdot \hat{\vec r}_2)\Big)},
    \label{eq: spin-spin}
\end{equation}
\c{where $\vec L$ is the orbital angular momentum of the system, and $\vec S_j$ and $\hat{\vec r}_j$ are, respectively the spin and the position unit vector of each black hole $j=1,2$. In addition,}
\begin{equation}
    \c{\vec S_{\rm eff}=\left(1+\frac{3m_2}{4m_1}\right)\vec S_1+\left(1+\frac{3m_1}{4m_2}\right)\vec S_2.}   
\end{equation}
A negative contribution in either term of the Hamiltonian would imply an ``attractive'' effective force\footnote{In the GR framework, ``repulsive'' or ``attractive'' effective forces mean divergent or convergent geodesics, respectively.}, and a positive contribution would yield a ``repulsive'' interaction. In terms of the spin-orbit interaction, the ``attraction'' occurs when the effective spin is anti-aligned with the angular momentum, and vice versa, while in terms of the spin-spin interaction, the ``attraction'' occurs when the spins are aligned with each other, and vice versa. This phenomenon is described in~\cite{Rettegno:2023ghr} for the spin-orbit interaction, but \c{we aim to study whether it becomes evident for spin induction on initially spinning black holes}.
\begin{figure}[tb!]
\centering
\includegraphics[width=\columnwidth]{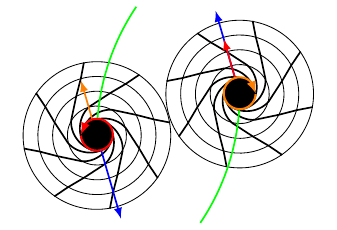}
\caption{Visualization of the spin induction mechanisms that appear when considering CHEs with spins along $\vec L$. The blue arrow represents the direction of the momentum during the CHE, while the other straight arrows represent the drag caused by the other spinning BH. The direction of the spin is presented as circular arrows. The frame-dragging is shown schematically, with the trajectory of each BH shown in green.}
\label{fig: Viz}
\end{figure}

\c{The system dynamics can be better interpreted by introducing the concept of precession, understood as the induced rotation on the corresponding inertial frame.} During a CHE, this phenomenon is observed with a certain precession velocity $\vec \Omega$ for each BH, which is then converted into spin due to conservation of angular momentum, as described for the non-spinning scenario~\cite{Jaraba:2021ces}. \c{However, the presence of an initial spin on the interacting black holes alters their dynamics. In particular}, a BH with negative spin causes frame-dragging against the precession direction of the other BH, while a BH with positive spin contributes to the precession direction of the other BH. Thus, when studying equally spinning BHs, the induced precession speed will be greater, increasing the spin-up with respect to the non-spinning scenario for aligned spins with $\vec L$, while it will decrease the spin induction for anti-aligned spins.  

On the other hand, when studying unequally spinning BHs, the BH with less initial spin will end up with a greater spin-up, because the one with the greater spin will contribute to a greater precession speed. \c{The opposite is true} for the BH with a greater initial spin, \c{as we depict} in Fig.~\ref{fig: Viz}. Additionally, there is a second-order effect (the spin-spin interaction), which \c{we also show} in Fig.~\ref{fig: Viz}, \c{due to which} both BHs try to repel each other, acting as a ``centrifugal'' force that increases $\vec\Omega$ when the difference between spins is increased. \c{This additional effect} will contribute to a greater spin-up on both BHs.

\c{The fact that spin induction can be understood as a combination of spin-orbit and spin-spin effects suggests it is possible to obtain analytical expressions for this effect, possibly in the parametrized post-Newtonian (PPN) formalism. As this article is a numerical exploration of spin induction with initial spins, we leave the computation of these expressions for future work.}


\section{Grid structure and initial conditions}  
\label{Sec: Grid}
The simulations were carried out using the Einstein Toolkit~\cite{Loffler:2011ay,EinsteinToolkit:web}. Specifically, the Cactus Computational Toolkit~\cite{Goodale:2002a,Cactuscode:web} was used, with the adaptive mesh refinement (AMR) provided by Carpet~\cite{Schnetter:2003rb,CarpetCode:web}, initial puncture data from TwoPunctures~\cite{Ansorg:2004ds,Paschalidis:2013oya} and the BSSN evolution from McLachlan~\cite{Brown:2008sb, Kranc:web, McLachlan:web}. Additionally, the AHFinderDirect thorn~\cite{Thornburg:2003sf,Thornburg:1995cp} was used to track the apparent horizons, QuasiLocalMeasures~\cite{Dreyer:2002mx} was used to measure the black hole spins and the Weyl scalar $\Psi_4$ was provided by the WeylScal4 thorn~\cite{Zilhao:2013hia}.

\begin{figure}
    \centering
    \includegraphics{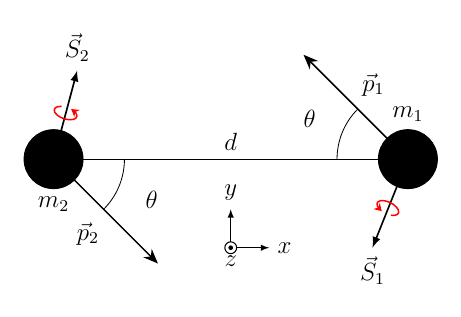}
    \caption{Initial conditions for the simulations. Here, $m_1= m_2$, $\vec{p}_1=-\vec{p}_2$. The spin components will generally be included along the $z$-axis, except in a specific case where we will describe the effects of spin perpendicular to $\vec L$.} 
    \label{fig: Cond-Ini}
\end{figure}

The initial configuration is identical to the one in~\cite{Jaraba:2021ces}, except for the inclusion of initial spins (see Fig.~\ref{fig: Cond-Ini}). \c{We only consider equal masses, $m_1=m_2\equiv m$, with symmetric initial momenta, $|\vec{p}_1|=|\vec{p}_2|\equiv p$. Since we work in geometrized units, the length, time, momenta, etc. have units of mass. As it is standard in Numerical Relativity, the total mass $M\equiv m_1+m_2=2m$ is fixed to 1 in the simulations, so for a generic value of total mass $M$, the results should be then rescaled by this quantity, leading us to present them as $t/M$, $p/M$, $x/M$, etc.}

We mainly study the effects of spins parallel to \c{$\vec L\equiv L\hat{z}$}, but we also describe some relevant effects that appear when considering spins perpendicular to $\vec L$. When analyzing the spin of a black hole, we define the dimensionless \c{vector} $\vec\chi=\vec S/m^2$, with \c{$0\leq\chi\leq1$}, $\chi=|\vec \chi|$, which will be used throughout the text.

We use the same grid properties as in~\cite{Jaraba:2021ces}, with half-lengths of $0.75\times 2^n$, for $n=0,1,\ldots,6,8,9,10$ and with steps $2^n\times\Delta x_{mr}$ for $n=0,1,\ldots,9$, considering $\Delta x_{mr}$ the size of the most refined grid. In our cases, we mainly use $\Delta x_{mr}= (3/200)M$ and $(3/256)M$ as our resolutions, which we respectively call \textit{medium} and \textit{high} resolutions, using the convention in~\cite{Jaraba:2021ces}. We also introduce a new resolution $\Delta x_{mr}= (1/104)M$, which we call \textit{very high} resolution and is used for some cases with particularly high spins that showed major differences between \textit{medium} and \textit{high} resolutions. These differences appear in some specific cases, \c{mainly} for spins \c{$\chi>0.6$}, \c{as} we will describe in the following sections.

\section{\c{Numerical results}}
\label{Sec: Results}
\c{We have run various simulations with equal masses, initial momenta $p/M=0.49$, and a selection of incidence angles $\theta\in[2.8^\circ, \ 3.03^\circ]$. While we aim to study purely hyperbolic encounters, the energy loss sometimes leads to both black holes becoming gravitationally bound, producing an eventual merger. These events are called dynamical captures (DC), producing two energy bursts: the hyperbolic-like part and the final merger. We try to avoid these events as much as possible, but the fact that we work with incidence angles close to the maximally spin-inducing case (which corresponds to the limit between CHE and DC), combined with the attractive effect due to the presence of initial spins (Sec.~\ref{Sec: Theory}), implies that some of our initial conditions fall into this regime. While there are no final spins to be measured in these cases, we extract the black hole spins between the hyperbolic-like interaction and the final merger, and include them in our plots, marking them clearly as non-scattering events.} 

Throughout this work, we use both notations $\chi_{\rm i}$ and $\chi(0)$ to determine the value of a spin variable at the start of a simulation.

\c{In this section, we study the precision of our simulations and then analyze results regarding equally spinning BHs. We leave the study of unequally spinning BHs for a separate section due to its importance and size.}
\subsection{Error analysis}
\label{subsec: error}

\c{In the case of equally spinning black holes (Subsec.~\ref{Subsec: Eq.theta, XEFF}), the differences between \textit{medium} and \textit{high} resolution are $\lesssim 2\%$ and $\lesssim 4\%$ for maximally spin-inducing scenarios and for lower spin-inducing scenarios, respectively. Therefore, we decided to use the \textit{medium} resolution for our simulations.} On the other hand, when studying unequally spinning BHs (Sec.~\ref{Sec: Numerical results 2}) we realised that higher resolutions are sometimes needed. For example, we observed that the \textit{medium} and \textit{high} resolutions presented higher discrepancies when $\chi_{\rm eff,i}\neq 0$, which may be due to additional recoil effects~\cite{Rettegno:2023ghr}. In fact, some specific cases may end up as DCs in the \textit{medium} resolution, while they are a CHE with \textit{high} resolution. Therefore, we decided to use \c{at least} the \textit{high} resolution for all cases involving unequally spinning BHs with $\chi_{\rm eff,i}\neq 0$, while keeping the \textit{medium} one for the other cases.

In any case, when studying unequally spinning BHs, we do not consider spins \c{$\chi>0.7$} because at least one of the BHs presents differences in the final spin $> 5\%$.

The plotted data points correspond to the maximum resolution used for each case. Furthermore, the errorbars of each point are obtained in two different ways:
\begin{itemize}
    \item When using the \textit{high} resolution, we compute the exact differences between the used resolution and the immediate lower resolution.
    \item When using the \textit{medium} resolution, we check the value with \textit{high} resolution for a few representative points, obtain the differences between resolutions, and do a regression for the rest of the data points.
\end{itemize}
\c{With this procedure, we} ensure the liability of the \c{results} while avoiding to use unnecessarily high resolution. 

In general, the differences between resolutions arise from the fact that a low resolution overestimates ``repulsive'' and ``attractive'' effective forces. This translates into an overestimation in the final spins for cases where attractive forces start to dominate with respect to lower initial spins. Conversely, spins tend to be underestimated in the cases where repulsive forces start to become more prominent with respect to lower initial spins. 

\subsection{Equally spinning black holes}
\label{Subsec: Eq.theta, XEFF}
First, we use the initial conditions $p/M=0.49$, $\theta=3.03^\circ$, a range of equal spins\footnote{Since the two BHs have the same initial spins and masses, the spin induction will be the same for both~\cite{Jaraba:2021ces}.} $\c{\chi_1^z(0)=\chi_2^z(0)=\chi^z}\in [-0.3,0.3]$ and \textit{medium} resolution for the data points in this section. Spins lower than $\chi^z=-0.3$ have not been considered because they produce dynamical captures for this choice of incidence angle $\theta$ and momentum $p/M$. Based on~\cite{Jaraba:2021ces,Rettegno:2023ghr}, we would expect that there is a noticeable spin induction and, due to additional ``repulsive'' forces for positive spins and ``attractive'' ones for negative spins (Eqs.~\ref{eq: spin-orbit} and~\ref{eq: spin-spin}), that the spin induction becomes less prominent \c{for larger} initial effective spins.

\begin{figure}[htb!]
    \centering
    \includegraphics[width=\columnwidth]{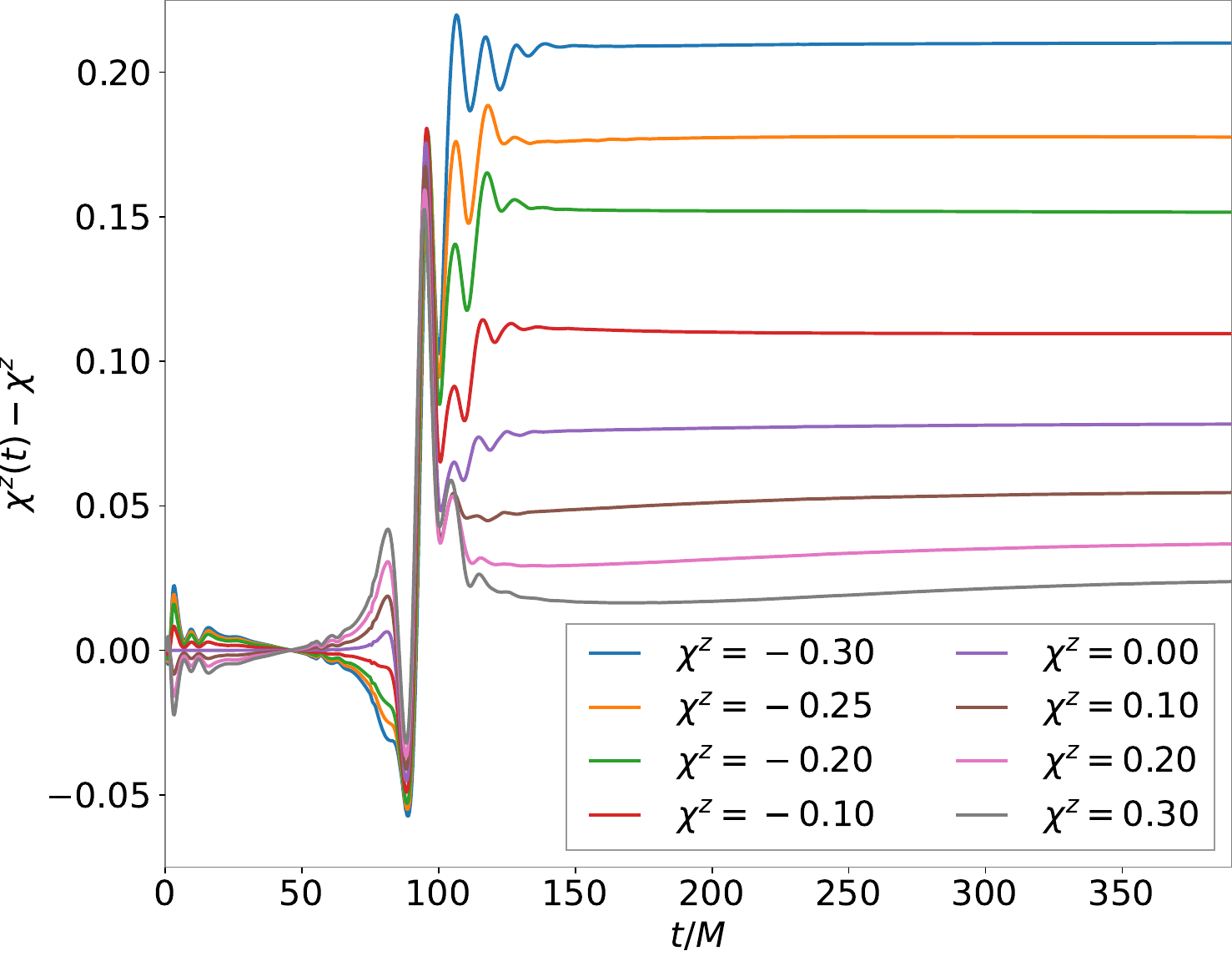}
    \caption{\c{Time evolution of the spin-up for 8 cases in the interval $\chi^z\in [-0.3,0.3]$}, with $p/M=0.49$ and $\theta=3.03^\circ$.}
    \label{fig: XEFF-various}
\end{figure}

This behavior is seen in Fig.~\ref{fig: XEFF-various}, where negative spins receive a greater spin induction than positive ones. Here, we can also observe that, for these initial conditions of $p/M$ and $\theta$, the spin-up in the lowest initial spin ($\chi^z=-0.3$) is $\sim 0.2$, which in the context of non-spinning cases can only be achieved for $p/M>0.7$~\cite{Jaraba:2021ces}. This shows that anti-aligned spins with respect to $\vec{L}$ suffer greater spin-up than aligned cases due to the new contribution in the Hamiltonian, as shown in Eq.~\eqref{eq: spin-orbit}, which also produces a closer hyperbolic encounter.

In general, when we include spin in our initial conditions, we initially observe some slight fluctuations in the \c{spin evolution due to the need of some relaxation time for the stabilization of the initial apparent horizons}. This is more noticeable for larger spins, but around $t/M=50$, the initial conditions have stabilized and, beyond this point, the spins are properly measured (see Fig.~\ref{fig: XEFF-various}).

\begin{figure}[htb!]
    \centering
    \includegraphics[width=\columnwidth]{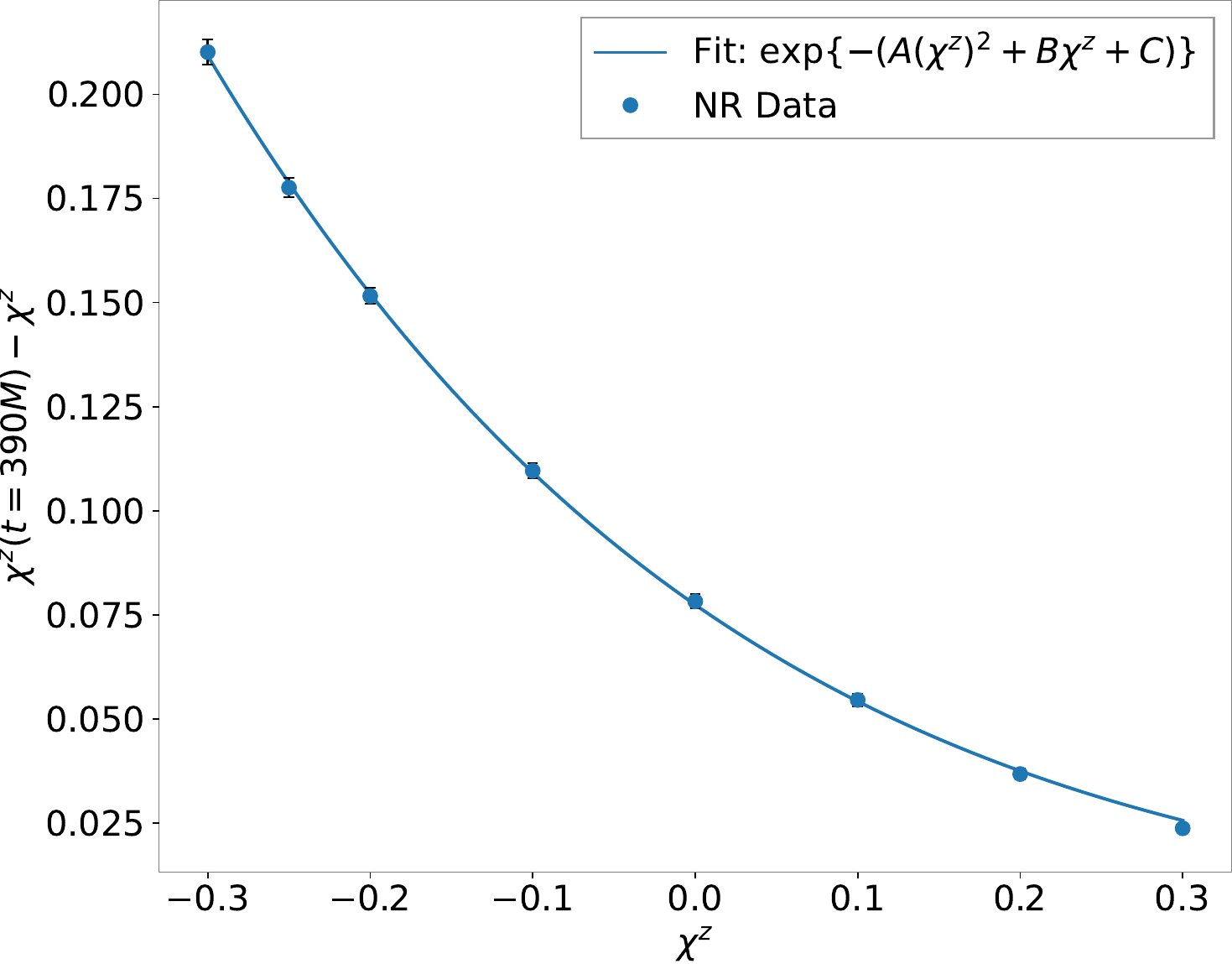}
    \caption{\c{Spin-up} at $t_f/M=390$ in terms of the initial spin for the cases represented in Fig.~\ref{fig: XEFF-various}, with an exponential fit.}
    \label{fig: XEFF-various-Fit}
\end{figure}

In Fig.~\ref{fig: XEFF-various-Fit}, we plot the final effective spin-up versus the initial effective spin. We observe that lower \c{$\chi^z$} induce more spin than higher ones, as \c{expected from} the Hamiltonian. \c{Additionally,} we fit our data to a decreasing exponential function, which seems to describe well the effective spin-up trend,
\begin{equation}
    \c{\chi^z(t_f)-\chi^z=\exp\{-(A(\chi^z)^2+B\chi^z+C)\}},
\end{equation}
with $A=0.61(17)$, $B=3.50(5)$ and $C=2.558(6)$, with 1$\sigma$ errors in parenthesis for the last digit(s).

\section{\c{Unequally} spinning black holes}
\label{Sec: Numerical results 2}
\c{We now consider different initial spins for each black hole.} In order to interpret the results, it is useful to define two spin-related variables\c{: the effective spin,}
\begin{equation}
    \c{\chi_{\rm eff}=\frac{\chi_1^z+\chi_2^z}{2},}
    \label{eq: chieff}
\end{equation}
\c{and the weighted spin difference,}
\begin{equation}
    \c{\delta\chi=\frac{\chi_1^z-\chi_2^z}{2}.}
    \label{eq: delchi}
\end{equation}
\c{The reason for these variable names is that, when spins are parallel to the orbital angular momentum $\vec{L}$, they coincide with the more general quantities defined for black-hole binaries~\cite{Gerosa_2023,Klein:2021jtd,Racine_2008,Damour_2001}},
\begin{align}
    \c{\chi_{\rm eff}}&=\c{\frac{\chi_1\cos\alpha_1+q\chi_2\cos\alpha_2}{1+q},}
    \label{eq: chieff gen} \\
    \c{\delta \chi}&\c{=\frac{\chi_1\cos\alpha_1-q\chi_2\cos\alpha_2}{1+q},}
    \label{eq: delchi gen}
\end{align}
\c{with $q=m_2/m_1$ and $\cos\alpha_{1,2}=\vec{L}\cdot\vec{S}_{1,2}/|\vec{L}||\vec{S}_{1,2}|$. In our case, as we work with equal masses, $q=1$ and, for spins parallel to $\vec{L}$, $\chi_j^z=\chi_j\cos\alpha_j$, $j=1,2$, thus reducing Eqs.~\eqref{eq: chieff gen} and~\eqref{eq: delchi gen} to Eqs.~\eqref{eq: chieff} and~\eqref{eq: delchi}. Although these parameters were originally introduced to describe black-hole binaries, their trends are much more clear than those of the individual spins (see Figs.~\ref{fig: eff-diff-fit}, \ref{fig: spinup-fit}). We thus adopt these variables to get a clearer understanding of the phenomenology.}

\c{In the following, we study four different scenarios}. Three of these have \c{$\vec S\parallel\vec L$}, with \c{$\chi_{\rm eff,i}\in\{0,-0.1,0.1\}$}, while the fourth one has \c{$\vec S\perp \vec L$}. This is \c{summarized} in Tab.~\ref{Tab: Method}, where we label and present the relevant information of \c{these scenarios, including their incidence angles, initial spin range and the minimum resolution considered, justified in Subsec.~\ref{subsec: error}}.

\c{In order to choose our incidence angles, we start from $\theta=2.86^\circ$, producing maximal spin induction in the non-spinning case.} Then, for each scenario, we \c{slightly vary this value} to compensate for the spin-orbit interaction (Eq.~\ref{eq: spin-orbit}) that appears when $\chi_{\rm eff,i}\neq 0$ \c{in the equally spinning case ($\chi_{1,p}(0)=\chi_{2,p}(0))$}, obtaining a spin-up similar to the non-spinning case. \c{The same value of $\theta$ is then used for all the unequally spinning cases corresponding to the same $\chi_{\rm eff,i}$.}

\begin{table}[t!]  
\caption{\c{Considered scenarios and their cases for the numerical simulations in Sec.~\ref{Sec: Numerical results 2}}, with $p/M=0.49$ and their respective \c{incidence angles $\theta$ and initial spins $\chi^k_{1,2}(0)$. The index $k$ corresponds to the axis along which the BHs are initially spinning: $k=z$ for scenarios a)-c) and $k=x$ for scenario d)}. The value $N_S$ represents the number of simulations that have been computed for the given scenario\c{, while} the ``Min. Res.'' is the minimum resolution considered for the plotted data points in this section.}
\centering 
\renewcommand{\arraystretch}{1.5}
\setlength{\tabcolsep}{3.3pt}
\begin{tabular}{|c|c|c|c|c|c|c|}
\hline\hline  
 Scenario & \c{Label} &$\theta$ & $N_S$ & \c{$\chi^k_{1,2}(0)$} & Min. Res. \\ [0.5ex]  
\hline   
a)  &$ \chi_{\rm eff,i}=0.0$ & $2.86^\circ$ & 7 & $[-0.6,0.6]$ & \textit{Medium}\\ [0.5ex] 
\hline   
b)  &$ \chi_{\rm eff,i}=-0.1$ & $2.92^\circ$ & 10 & $ [-0.7,0.5]$ & \textit{High} \\ [0.5ex] 
\hline   
c)  &$ \chi_{\rm eff,i}=0.1$ & $2.80^\circ$ & 9 & $ [-0.5,0.7]$ & \textit{High} \\ [0.5ex]
\hline
d)  &$\vec S\perp\vec L$ & $2.86^\circ$ & 6 & $ [-0.4,0.4]$ & \textit{Medium} \\
\hline
\hline
\end{tabular}  
\label{Tab: Method}
\end{table} 

For the three scenarios with \c{$\vec S\parallel\vec L$}, we analyze the \c{absolute change (spin-up) in the} variables $\chi_{\rm eff}$, $\delta\chi$ and \c{$\chi_{1,2}^z$} at a \c{certain final time $t_f$. We use $t_f=285M$, which we found to be long enough for the spins to stabilize after the strong interaction}. The first two variables are \c{fitted to the data using simple parametrizations. For the \c{$\vec S\perp\vec L$} case,} on the other hand, we will plot the spin-up for each \c{spin component}.

\subsection{Spins parallel to $\vec L$}
\label{Subsec: figs, analysis}
\begin{figure*}[hbt!]
\centering
  \includegraphics[width=\textwidth]{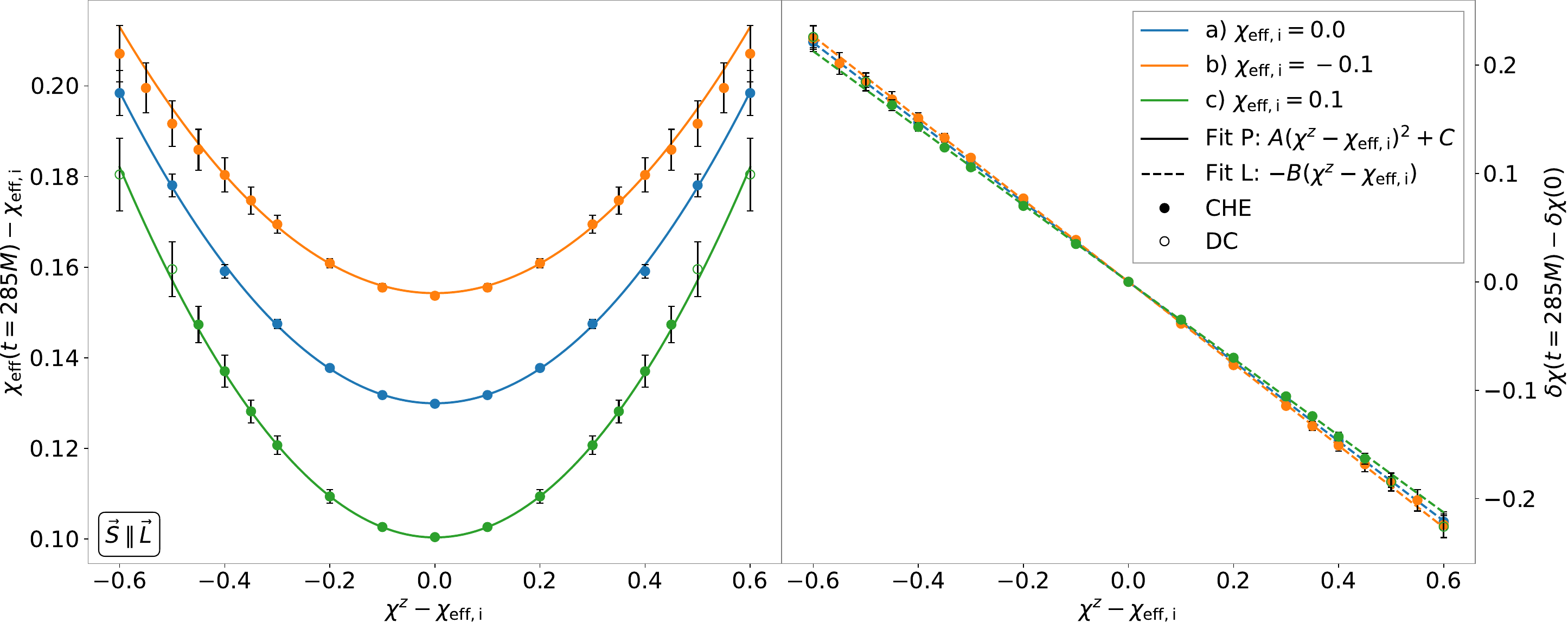}
  \caption{\c{Effective spin (left) and weighed spin difference change (right) values at $t_f/M=285$ along with their respective fits. The fitted coefficients ($A,\ B,\ C$) along with their uncertainties, are shown in Tab.~\ref{Tab: Fit}.}}
  \label{fig: eff-diff-fit}
\end{figure*}
\begin{figure}[hbt!]
\centering
  \includegraphics[width=\columnwidth]{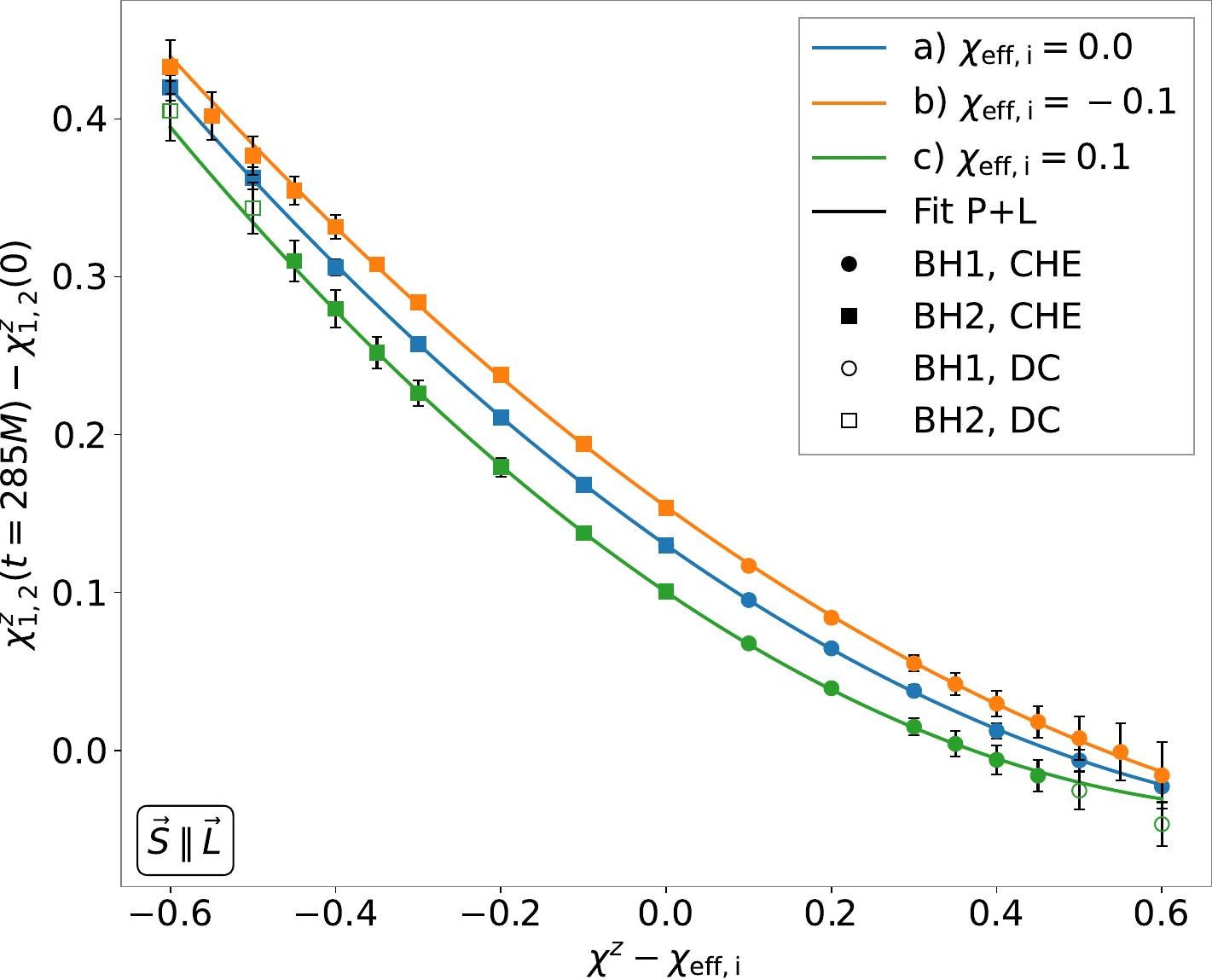}
  \caption{\c{Spin-up at $t_f/M=285$ for each BH with the fits computed in Fig.~\ref{fig: eff-diff-fit}, with the coefficients shown in Tab.~\ref{Tab: Fit}. }}
  \label{fig: spinup-fit}
\end{figure}

In Fig.~\ref{fig: eff-diff-fit}, \c{we plot the spin-up on $\chi_{\rm eff}$ (left) and on $\delta\chi$ (right) for the three scenarios of BHs spinning along $\vec L$, a)-c). On the horizontal axis, we set the $z$ component of the first BH's initial spin, shifted by the initial $\chi_{\rm eff}$ so as to have the three scenarios horizontally centered at the same point. Notice that it is only necessary to run the simulations corresponding to half of the horizontal axis, as the other half is recovered by symmetry: interchanging both black holes leaves $\chi_{\rm eff}$ invariant (even parity), while it changes the sign of $\delta\chi$ (odd parity), as deduced from the definitions in Eqs.~\eqref{eq: chieff} and~\eqref{eq: delchi}.}

\c{On the left panel of Fig.~\ref{fig: eff-diff-fit},} we observe that a larger \c{spin asymmetry enhances the effective} spin induction, due to additional spin-orbit and spin-spin interactions that are not considered in the study of non-spinning CHEs~\cite{Jaraba:2021ces}. The trend of this plot can be described by a simple parabola,
\begin{equation}
    \c{\chi_{\rm eff}(t_f)-\chi_{\rm eff,i}=A(\chi^z-\chi_{\rm eff,i})^2+C,}
    \label{eq: chieff fit}
\end{equation}
\c{where the fitted coefficients are provided in Tab.~\ref{Tab: Fit} for each scenario, together with their 68\% confidence-level errors. As shown in Fig.~\ref{fig: eff-diff-fit}, two simulations with $\chi_{\rm eff, i}$ are actually dynamical captures, which we do not consider for our fits, as they are not hyperbolic events. Nevertheless, as they have not reached the merger phase by $t=t_f$, we plot their spins measured at this time, confirming they still follow the observed trends, even if some deviation was to be expected.}

The remarkable accuracy of these fits, even for initial spins \c{$\chi>0.5$}, may hint at the existence of an analytical explanation to this phenomenon. We observe, however, that the \c{effective spin} curve corresponding to scenario b) does not fit equally well \c{the three last points on both sides of} the parabola. \c{This may be due to} the big differences between \textit{medium} and \textit{high} resolution for these points, \c{visible in the errorbars, which suggests the need to use higher resolutions for more asymmetric cases}.

\begin{table}[b!]  
\caption{Fitted parameters at 68\% confidence level for \c{$\chi_{\rm eff}(t_f)-\chi_{\rm eff,i}$} and $\delta\chi(t_f)-\delta\chi(0)$ in terms of initial spins in Fig.~\ref{fig: eff-diff-fit}, for the first 3 scenarios, a)-c). The errors at $1\sigma$ are indicated in brackets for the last digit.}
\centering
\renewcommand{\arraystretch}{1.5}
\setlength{\tabcolsep}{13.5pt}
\begin{tabular}{|r|c|c|c|}
\hline\hline  
 $\chi_{\rm eff,i}$ & $A$ & $B$ & $C$ \\ [0.5ex]  
\hline
$0.0$ & $0.1909(13)$ & $0.3676(3)$ & $0.1300(1)$ \\ [0.5ex] 
\hline   
$-0.1$ & $0.1631(27)$ & $0.3776(8)$ & $\c{0.1542}(4)$ \\ [0.5ex] 
\hline   
$0.1$ & $0.2269(17)$ & $0.3546(5)$ & $\c{0.1003}(2)$ \\
\hline
\hline
\end{tabular}  
\label{Tab: Fit}
\end{table} 

On the other hand, we can carry out a similar analysis on $\delta\chi(t_f)-\delta\chi(0)$, which is shown \c{on the right panel of Fig.~\ref{fig: eff-diff-fit}}. Here, we observe \c{that} the weighed spin difference, subtracted its initial value, decreases for higher \c{values of $\chi^z$}, following an \c{approximately} linear trend. The fit for this variable then follows the equation
\begin{equation}
    \c{\delta\chi(t_f)-\delta\chi(0)=-B(\chi^z-\chi_{\rm eff,i}),}
    \label{eq: delchi fit}
\end{equation}
\c{where the fitted coefficients (again, excluding DCs) are also provided in Tab.~\ref{Tab: Fit}.} 

\c{We observe that, as the value of $\chi_{\rm eff,i}$ increases, the parabolae are more pronounced, with lower spin-ups. Additionally, the values of $B$ are similar in the three scenarios, with the largest difference being $<7\%$, showing that this coefficient does not drastically vary with $\chi_{\rm eff,i}$.} 

\c{Going back to the individual spins using Eqs.~\eqref{eq: chieff} and~\eqref{eq: delchi}, the approximations in Eqs.~\eqref{eq: chieff fit} and~\eqref{eq: delchi fit} imply}
\begin{equation}
    \c{\chi_1^z-\chi_1^z(0)}=A(\c{\chi^z}-\chi_{\rm eff,i})^2\c{-}B(\c{\chi^z}-\chi_{\rm eff,i})+C,
    \label{eq: chi1}
\end{equation}
\begin{equation}
    \c{\chi_2^z-\chi_2^z(0)}=A(\c{\chi^z}-\chi_{\rm eff,i})^2\c{+}B(\c{\chi^z}-\chi_{\rm eff,i})+C.
    \label{eq: chi2}
\end{equation}

\c{Using} these equations, we plot the final spin-up of each BH \c{and their fits} in Fig.~\ref{fig: spinup-fit}. Here, we can see how the spin-up of each BH drifts from the equally spinning case \c{as the initial spins become more important}.

These results match \c{the expected behaviour described} in Sec.~\ref{Sec: Theory}, where we showed that, at first order, the BH with lower initial spin \c{acquires} a greater spin-up. At second order, \c{the spin-spin contribution enhances the spin-up for both BHs as their initial spin asymmetry increases, producing} the parabolic trend.

Moreover, \c{certain points corresponding to some of the highest initial spins} in Fig.~\ref{fig: spinup-fit} \c{show} negative spin-ups. This can be \c{interpreted as an effect of the negatively spinning BH, which, in this cases, dominates over the other interaction terms in Eq.~\eqref{eq: hamiltonian}. In particular, this implies that spin-spin interactions govern the spin dynamics of these CHEs.}

\begin{figure*}[hbt!]
\centering
  \includegraphics[width=\textwidth]{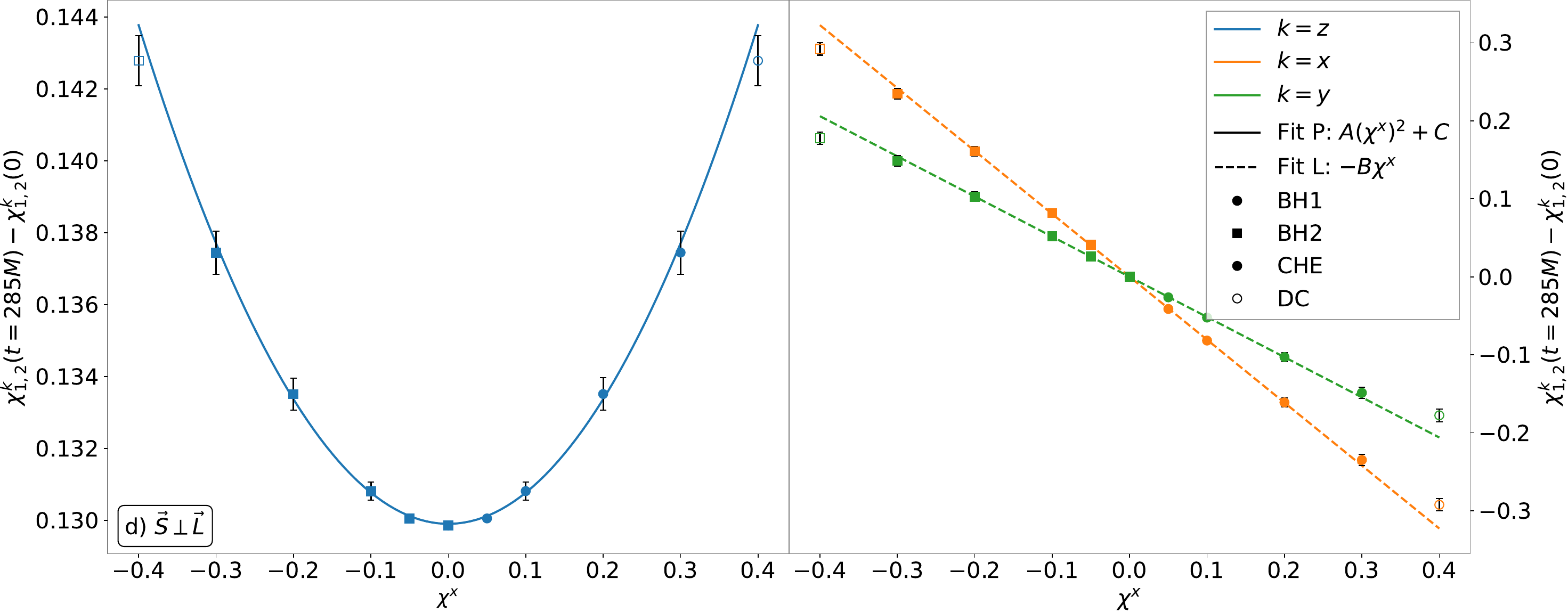}
  \caption{Spin-up in the $z$ direction (left) and $x,\ y$ directions (right). The fitted coefficients ($A,\ B,\ C$) along with their uncertainty are shown in Tab.~\ref{Tab: XFit}.}
  \label{fig: XYZ-fit}
\end{figure*}
\subsection{Spins orthogonal to $\vec L$}
\c{We conclude our numerical exploration with a series of simulations with BH initial spins orthogonal to $\vec{L}$ (scenario d)). The initial conditions in spin are chosen so that $\chi^x_1(0)=-\chi^x_2(0)=\chi^x$, with the rest of the components set to zero. Due to the symmetries in the system, interchanging the BH spins leaves the $z$ component invariant, but changes the sign of the $x$ and $y$ components.}

\c{In Fig.~\ref{fig: XYZ-fit}, we plot the spin-up on each spin component. On the left panel, we observe that spin induction on the $z$-axis has a parabolic dependence on the initial spin}, which hints \c{that} spin-spin interaction \c{may be the cause of this behaviour}. In Eq.~\eqref{eq: spin-spin}, we see that there is indeed a coupling of the interaction for spinning BHs orthogonal to $\vec L$.

Furthermore, \c{given that} the spin-orbit term (Eq.~\eqref{eq: spin-orbit}) \c{now} cancels out \c{initially}, the spin-spin term (Eq.~\eqref{eq: spin-spin}) plays a \c{more relevant} role, with this Hamiltonian being more attractive \c{for larger} $|\chi^x|$. \c{In addition,} similarly \c{to} the scenarios a)-c), the \c{spin-up of the $z$ component is enhanced for larger differences between the initial spins}.

\c{On the right panel of} Fig.~\ref{fig: XYZ-fit}, we see that \c{the spin-up in both directions $x$ and $y$ has an approximately linear dependence on $\chi^x$. This behavior can be explained from the fact that, when spins are not parallel to the $z$-axis, they precess, understanding precession as a temporal evolution of the form}
\begin{equation}
   \c{ \frac{\mathrm{d} \vec S}{\mathrm{d}t}=\vec\Omega\times\vec S},
    \label{eq: Precession}
\end{equation}
\c{for a certain precession vector $\vec{\Omega}$ which, due to the symmetry of the system, is the same for both BHs and parallel to $\vec L$.} 

\begin{figure}[ht!]
\centering
\includegraphics[width=\columnwidth]{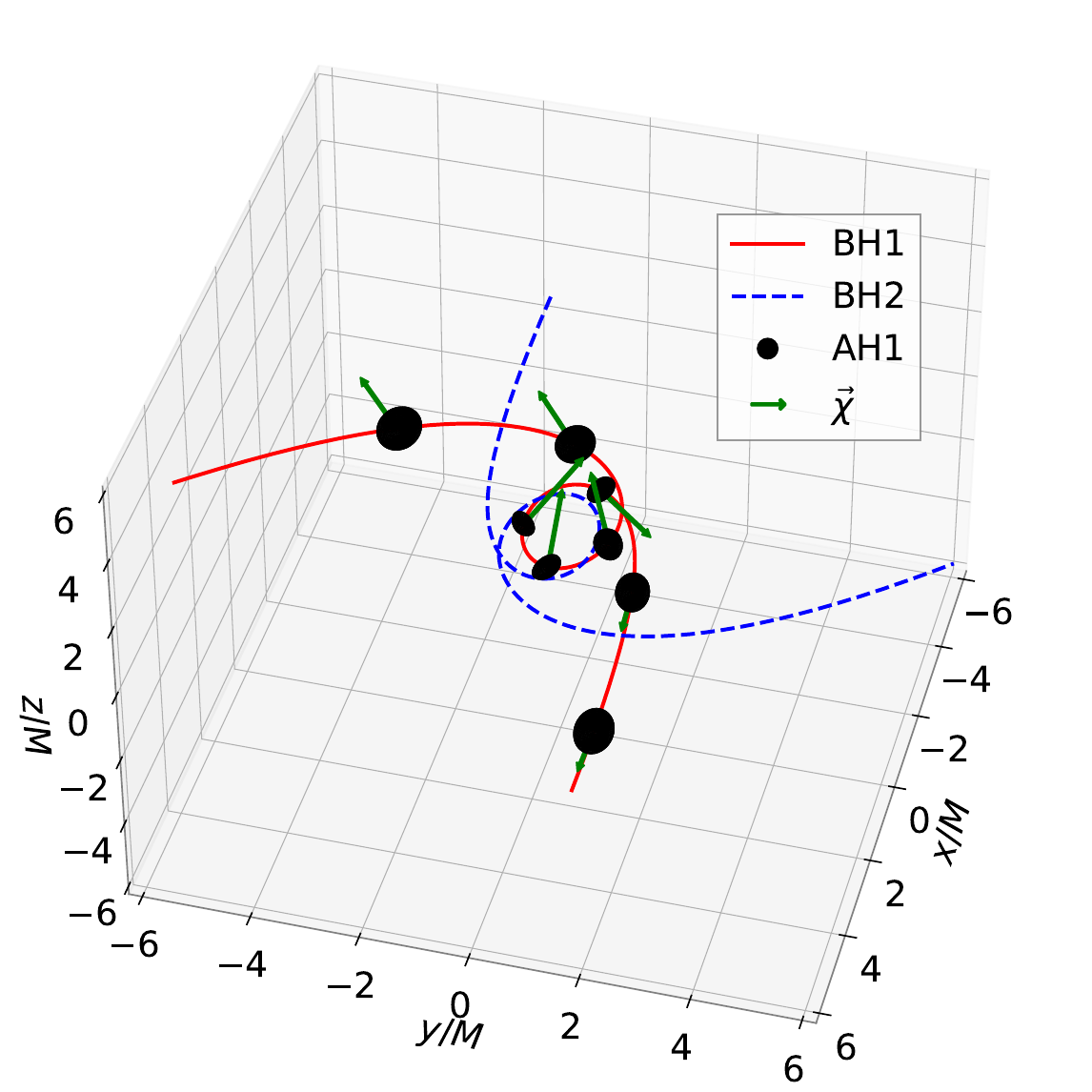}
\caption{Visualization of the precession of the BH spins due to a CHE with initial spin pointing in the $x$-axis and facing opposite directions.}
\label{fig: Precc}
\end{figure}

\c{In Fig.~\ref{fig: Precc}, we plot one of the trajectories of scenario d) in three dimensions, together with the spin vectors, in order to illustrate the effects of precession. We observe how the spin vector changes direction following the direction of $\vec L$, following Eq.~\eqref{eq: Precession}.}
\c{Here, we also see that, during this precession, the black holes receive a boost in the $z$ direction towards positive $z$, while exchanging the BH spins would cause motion towards $z<0$. In either case, the direction of the orbital plane does not change, as the total spin is zero, thus not inducing any precession on $\vec{L}$.}

Finally, \c{the spin components can be approximated as}
\begin{align}
    \c{\chi_{1,2}^{k}(t_f)-\chi^{k}_{1,2}(0)}&\c{=\mp B^{k}\chi^x, \ k=x,y;}\\
    \c{\chi_{1,2}^z(t_f)-\chi^z_{1,2}(0)}&\c{=A(\chi^x)^2+C}.
\end{align}
These fitted coefficients are presented in Tab.~\ref{Tab: XFit}, where we also show the 68\% confidence-level errors. \c{We exclude the dynamical captures ($\c{\chi^x}=\pm 0.4$), as done in the previous subsection, observing that they do not fit very well the linear trend from the other points.}

\begin{table}[hb!]  
\caption{Fitted parameters at 68\% confidence level for $\chi_{1,2}^k(t_f)-\chi_{1,2}^k(0)$, with $k=x,\ y,\ z$ in terms of the initial spin $\chi^x$. We show the errors in parentheses at $1\sigma$.}  
\centering 
\renewcommand{\arraystretch}{1.5}
\setlength{\tabcolsep}{9pt}
\begin{tabular}{|r|c|c|c|c|} 
\hline\hline  
 Axis & Fit & $A$ & $B$ & $C$ \\ [0.5ex]  
\hline   
$x$ & L & -- & $0.8067(28)$ & -- \\ [0.5ex]
\hline   
 $y$ & L & -- & $0.5150(20)$ & -- \\ [0.5ex]
\hline   
$z$ & P& $0.0867(13)$ & -- & $0.1300(1)$ \\ [0.5ex] 
\hline 
\hline
\end{tabular}  
\label{Tab: XFit}
\end{table} 

\subsection{Trajectories and spin-up efficiency}
\begin{figure*}[hbt!]
\centering
  \includegraphics[width=\textwidth]{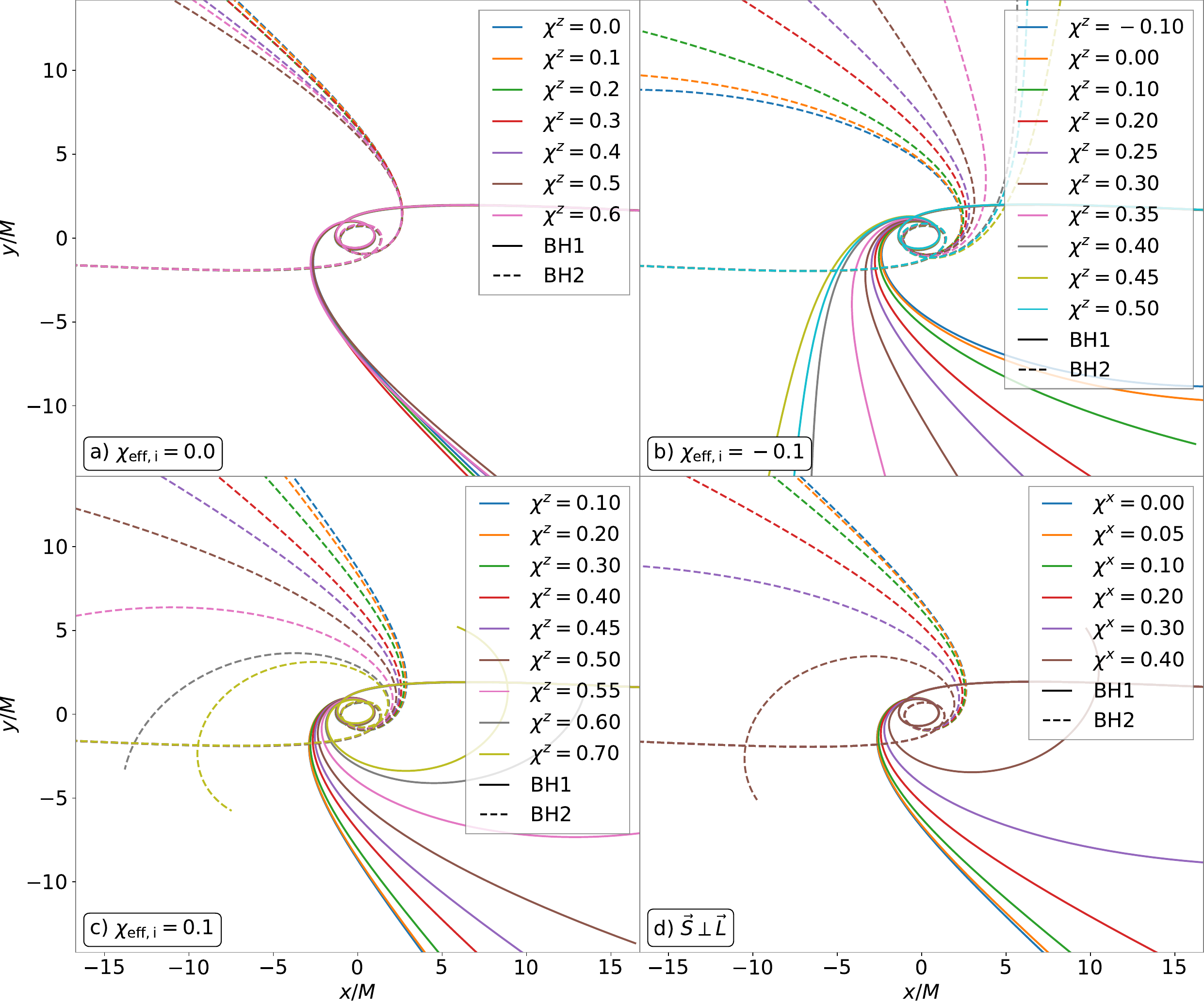}
  \caption{Trajectories of each CHE considered in Sec.~\ref{Sec: Numerical results 2}. A negative effective spin seems to create more repulsive CHEs for bigger spin differences, while the opposite occurs for positive effective spins, as seen in scenarios b) and c), respectively. Zero initial effective spin does not have great effects on the scattering angles (scenario a)). Varying the spins in the $x$ direction tends rapidly to the dynamical capture (scenario d)).}
  \label{fig: tray}
\end{figure*}

\c{In order to better understand the behavior of the simulations, we plot their} trajectories in Fig.~\ref{fig: tray}. We observe that the scattering angles change when varying the BH spins, \c{up to the point of producing the dynamical captures mentioned before}. We can also see that the cases on scenario a) suffer less differences between the final trajectories when varying the spins, while the other three scenarios have greater differences in terms of scattering angles. Here, we can also observe that scenario b) is more repulsive \c{as the spin difference increases}, while scenarios c) and  d), \c{on the contrary, produce more attraction for larger spin asymmetry. This phenomenology is hard to understand just from the spin-spin term in Eq.~\eqref{eq: spin-spin}, which motivates further investigation on these effects}.

\c{Moreover}, the scenario d) seems to \c{generate closer encounters} (see Fig.~\ref{fig: tray}) for relatively low spins ($|\chi^x|\gtrsim0.3$), \c{which may be due to the fact that frame-dragging caused by opposite spins in the $x$ axis pulls both BHs closer when the boost in the $z$ axis is induced, as earlier seen in Fig.~\ref{fig: Precc}}. In fact, greater values of $\chi^x$ have not been included since they \c{led to} dynamical captures \c{at} $t/M<300$. This ``attractive'' effect is also \c{expected from} the spin-spin Hamiltonian in Eq.~\eqref{eq: spin-spin}. 

Furthermore, it is worth mentioning that we \c{did not consider to incorporate a rigorous study} of scattering angles because we \c{obtained} errors up to $\sim8^\circ$. \c{Higher resolutions would thus be required for a more detailed discussion on scattering angles}. 

\begin{figure}[ht!]
\centering
\includegraphics[width=\columnwidth]{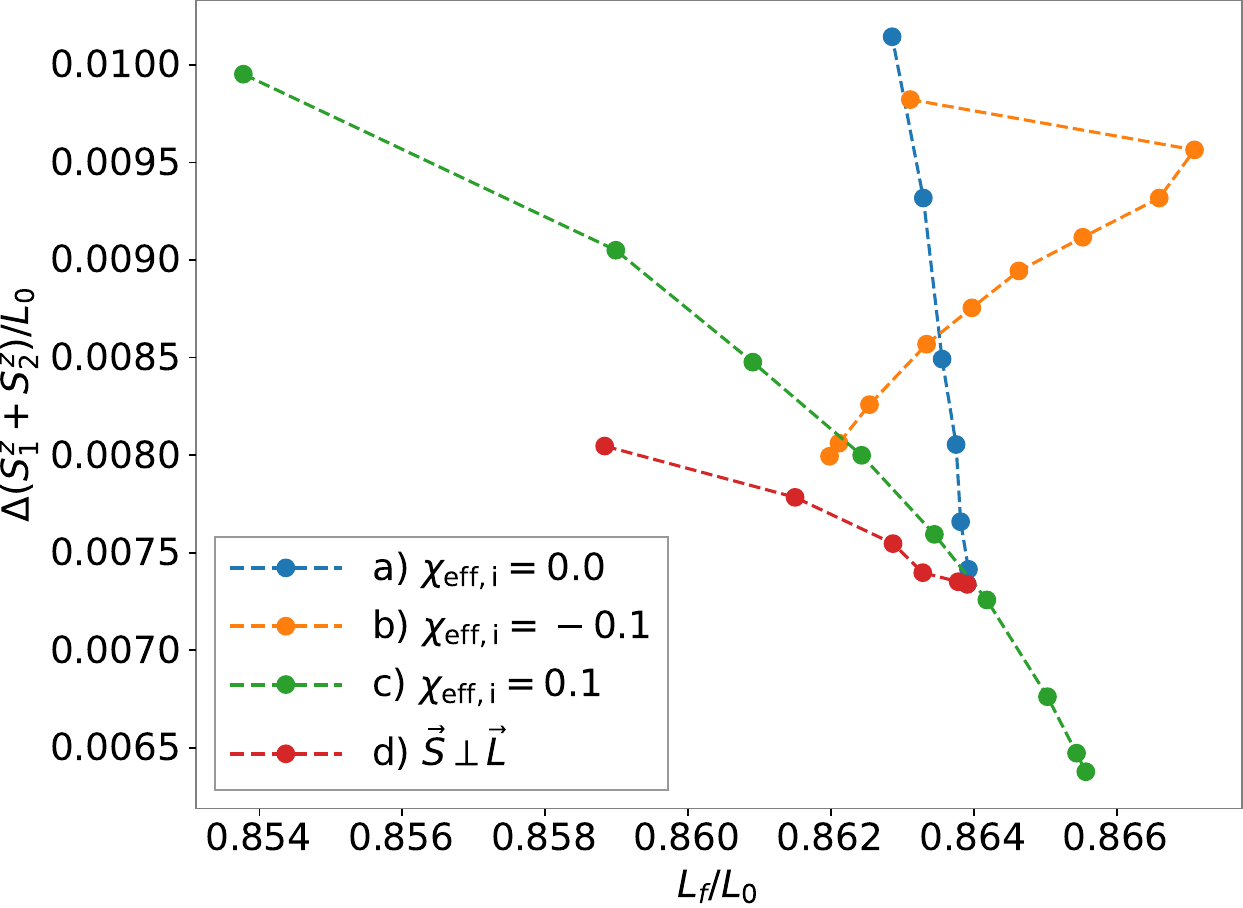}
\caption{\c{Total} spin-up weighed with the initial angular momentum with respect to the final fraction of angular momentum for each scenario.}
\label{fig: SUE}
\end{figure}

\c{Finally, we show} the spin-up efficiency in Fig.~\ref{fig: SUE}, \c{adapting} the procedure in~\cite{Nelson:2019czq} \c{to the presence of initial spins. \c{We plot the ratio between} the total spin-up $\Delta (S_1^z+S_2^z)=S^z_1(t_f)+S_2^z(t_f)-S^z_1(0)-S_2^z(0)$ and the orbital angular momentum ($L_0$) with respect to the remaining fraction of orbital angular momentum ($L_f/L_0$) after the CHE, with $L_f=L_0-\Delta (S_1^z+S_2^z)-L_{\rm GW}$. Our results agree with those in}~\cite{Nelson:2019czq} and~\cite{Chiaramello:2024unv} for this initial boost \c{$p/M=0.49$, taking into account} that our initial conditions generate slightly closer hyperbolic encounters. \c{The correlation} with the scattering angles \c{in} Fig.~\ref{fig: tray} \c{is more obvious here}, since the GW emission is greater for more closely scattered BHs, as expected. We also see that we can achieve a spin-up efficiency $\gtrsim1\%$ for the highest spins with this initial \c{momentum}.

\c{We observe that negative effective spins cause $L_f/L_0$ to grow with the spin induction, obtaining a positive slope, while \c{the opposite behavior occurs in} the other scenarios. However,} the most efficient point for scenario b) creates a discontinuity in the curve. This is due to the scattering angle measured for this CHE, which seems to break the tendency seen in Fig.~\ref{fig: tray}. \c{This may hint at} the need for a higher precision for such highly spinning BHs with the given initial conditions.

With this, we see that the \c{inclusion of initial spins} creates \c{more complex} trends in the \c{spin efficiency plot, which were not present in the case of initially non-spinning BHs~\cite{Nelson:2019czq}}.

\c{In general, initial spins notably} affect the spin induction mechanisms \c{studied in previous works~\cite{Nelson:2019czq,Jaraba:2021ces}}, their trajectories and their GW emission, \c{showing} how rich and complex the phenomenology of CHEs truly is.

\section{Conclusions}
\label{Sec: Concl}
In this article, we have studied the spin induction effect that arises in both black holes involved in a close hyperbolic encounter. We have done this by running numerical relativity simulations, following  Refs.~\cite{Nelson:2019czq,Jaraba:2021ces}, with the introduction of initial spins for the black holes.

First, we have considered a series of simulations in which the initial spins were set to be equal in both black holes, and aligned with the orbital angular momentum. In this case, we have shown how the spin-up becomes weaker for larger initial spins, due to the spin-orbit interaction, Eq.~\eqref{eq: spin-orbit}. This trend is well approximated by an exponentially decreasing function, as shown in Fig.~\ref{fig: XEFF-various-Fit}.
We have then considered other scenarios in which the initial spins are unequal, including aligned and orthogonal spins with respect to the orbital angular momentum. We observed that, given a fixed initial effective spin, the induced effective spin becomes larger for greater values of $\chi_1(0)$ and $\chi_2(0)$. This behaviour can be modelled by a second-order polynomial, both for the aligned cases and for the $z$ component of the orthogonal cases, as can be seen in Fig.~\ref{fig: eff-diff-fit} and~\ref{fig: XYZ-fit}. Unlike in the equally spinning case, there is now a non-zero $\delta\chi$, which follows an approximately linear trend, as shown in Fig.~\ref{fig: eff-diff-fit}. Combining both equations, we provide effective quadratic expressions for $\c{\chi_1^z-\chi_1^z(0)}$ and $\c{\chi_2^z-\chi_2^z(0)}$, see Fig.~\ref{fig: spinup-fit}. An intuitive explanation of this phenomenon is shown in Fig.~\ref{fig: Viz}. We have shown as well that the final spin-up in the $x$ and $y$ components, result of the precession of the spins (see Fig.~\ref{fig: Precc}), as defined by Eq.~\eqref{eq: Precession}, follows an approximately linear trend as well, see Fig.~\ref{fig: XYZ-fit}.

We have also shown that the inclusion of spins in our initial conditions has relevant implications on the trajectories (scattering angles) and emission of GWs of the CHEs, see Figs.~\ref{fig: tray} and~\ref{fig: SUE}. This may be relevant for analytical studies of the effects discussed throughout this work, as done in~\cite{Chiaramello:2024unv} for the previous article on spin induction~\cite{Jaraba:2021ces}. 

An accurate modelling of the spin induction effect is necessary for a good understanding of the clustering properties of black holes, which has drastic implications, in particular, for primordial black holes. It is expected that CHEs in dense black hole clusters are relevant for the spin distribution of the BH population~\cite{Garcia-Bellido:2020pwq}, which is why it is essential to not only consider the study of spinless scattering~\cite{Jaraba:2021ces, Nelson:2019czq} but include spins parallel and orthogonal to $\vec L$, as done in this work. 

Numerical works as this one progressively contribute to our understanding of this phenomenon, introducing diverse perspectives as different parameter areas are explored. Further exploration of the parameter space is thus needed, for instance, with the study of initially spinning black holes of unequal masses, which is left for future work. In any case, an accurate and complete description of this effect will require analytical studies, making use of post-Newtonian, post-Minkowskian, effective one-body or similar approaches. In the absence of these analytical studies, however, NR simulations remain a very powerful tool to explore the phenomenology of spin induction.

\begin{acknowledgments}
All the simulations have been run in the Hydra HPC cluster at the IFT. SJ and JGB acknowledge support from the Spanish Research Project PID2021-123012NB-C43 [MICINN-FEDER], and the Centro de Excelencia Severo Ochoa Program CEX2020-001007-S at IFT. SJ is supported by the FPI grant PRE2019-088741 funded by MCIN/AEI/10.13039/501100011033.
\end{acknowledgments}

\bibliography{main}
\end{document}